\documentclass[10pt,a4paper,oneside]{article}
\usepackage{graphicx}
\usepackage{color}
\usepackage{booktabs}
\usepackage{epsfig}
\usepackage{multirow}
\usepackage[english]{babel}
\usepackage{amsmath}
\usepackage{amssymb}
\usepackage[latin1]{inputenc}
\usepackage{fancyhdr}
\usepackage{indentfirst}
\usepackage{newlfont}
\usepackage{latexsym}
\usepackage[latin1]{inputenc}
\usepackage{fancyhdr}
\usepackage{booktabs}
\usepackage[hang,bf,small]{caption}
\usepackage{microtype}
\usepackage[a4paper,top=4cm,left=4cm]{geometry}
\title{Haantjes Manifolds and Veselov Systems} 
\author{F. Magri\\
\normalsize{Dipartimento di Matematica e Applicazioni}\\
\normalsize{Università di Milano Bicocca}\\
\normalsize{franco.magri$@$unimib.it }\\
}
\date{}
\begin{document}
\maketitle 
\begin{abstract}
A novel geometric interpretation of the solutions of the WDVV equations formerly found by A.P. Veselov is suggested. 
\end{abstract}
\section{Introduction}
In 1999 A.P. Veselov discovered a class of special solutions of the Witten-Dijkgraaf-Verlinde-Verlinde equations, playing an important role in 2D topological field theory and N=2 SUSY Yang-Mills theory. His solutions are related to special collections of covectors in a linear space called V-systems $\cite{1},\cite{2}$. A representative example of this class of solutions is the function
\[ F(x_1,...,x_n)=\sum_{i<j}^n (x_i-x_j)^2\log(x_i-x_j)^2+\frac{1}{m}\sum_{i=1}^nx_i^2\log x_i^2 \] 
depending on an arbitrary parameter m. The corresponding V-system is a deformation of the root system $A_n$.

More recently I have proposed to consider special arrangements of 1-forms on a manifold called Lenard complexes $\cite{3}$. In this paper I argue that the V-system of Veselov may be related to a particular class of Lenard complexes, qualified as equivariant complexes.

My argument proceeds in four steps. In Sec.2 I recall the definition of Lenard complex and I give two examples. In Sec.3 I explain the relation between the solutions of the WDVV equations and the Lenard complexes. One complex of this type is associated with any solution of the WDVV equations, and viceversa. In Sec.4 I introduce the notion of equivariant Lenard complex. In Sec.5, finally, I study a particular class of equivariant complexes which seems to be related to the V-systems of Veselov.
\section{Lenard complexes}
A Lenard complex is a composite geometric structure which may be defined over a special class of manifolds called Haantjes manifolds. It may be viewed as a mild generalization of the Lenard chains originally defined on bihamiltonian manifolds.

To define a Lenard complex on a manifold M, one needs:
\begin{enumerate}
\item \begin{tabular}{p{0.5\columnwidth}@{\quad}l}
a vector field  & $X:M \rightarrow TM$ 
\end{tabular}
\item \begin{tabular}{p{0.5\columnwidth}@{\quad}l}
an exact 1-form  & $dA:M \rightarrow T^*M$ 
\end{tabular}
\item {\begin{tabular}{p{0.5\columnwidth}@{\quad}l}
a family of tensor fields of type (1,1) & $K_j:TM \rightarrow TM$
\end{tabular}}
\end{enumerate}
in number equal to the dimension of the manifold. According to a consolidated tradition, these tensor fields will be referred to as the ``recursion operators'' of the complex. By assumption, they pairwise commute:
\renewcommand\theequation{\Roman{equation}}
\begin{equation}
K_jK_l-K_lK_j=0.
\end{equation}
Their action on X and dA gives rise to the usual chains of vector fields
\[ X_j=K_jX, \]
and of 1-forms
\[ \theta_j=K_jdA. \]
More importantly they also give rise to the (symmetric) square of 1-forms
\[ \theta_{jl}=K_jK_ldA. \]
This square of forms is the main novelty of the theory of Lenard complexes with respect to the old theory of Lenard chains. Another difference is that the recursion operators $K_j$ are not the powers of a single operator K. A third difference is that nothing is assumed, a priori, on the torsion of the recursion operators $K_j$. They may have torsion. Notwithstanding, it can be shown that the recursion operators of a Lenard complex have always vanishing Haantjes torsion. This is the ultimate reason to call Haantjes manifolds the manifolds supporting a Lenard complex.

\bigskip
\noindent
\textbf{Definition 1.}\textit{The composite structure formed by the chain of vector fields $X_j$, by the chain of 1-forms $\theta_j$, and by the square of 1-forms $\theta_{jl}$ is a Lenard complex on the Haantjes manifold M if
\begin{equation}
[X_j,X_l]=0
\end{equation}
\begin{equation}
d\theta_j=0
\end{equation}
\begin{equation}
d\theta_{jl}=0,
\end{equation}
that is if the vector fields commute and if the 1-forms both of the chain and of the square are closed and, therefore, locally exact.}

\bigskip
\noindent
If furthermore the vector fields $X_j$ and the 1-forms $\theta_l$ are linearly independent (in some open dense set of M), the previous assumptions entail that a Lenard complex define two distinct coordinates systems on the base manifold $M$.

\bigskip
\noindent
\textbf{Definition 2.}\textit{ The coordinates $x_j$ and $a_j$ locally defined by the relations
\[ K_jX=\frac{\partial}{\partial x_j} \] 
\[ K_jdA=da_j\] 
are called the ``x-coordinates'' and the ``a-coordinates'' induced by the Lenard complex on the Haantjes manifold M.}

\bigskip
In addition to the coordinates $x_j$ and $a_j$, it is useful to keep trace also of the local potentials of the closed 1-forms $\theta_{jl}$.

\bigskip
\noindent
\textbf{Definition 3.} \textit{The symmetric matrix whose entries are the local potentials $A_{jl}$ of the 1-forms of the Lenard square 
\[ K_jK_ldA=dA_{jl}, \]
is called the matrix potential of the Lenard complex.}

\bigskip
Nothing else is really required to proceed rapidly towards the WDVV equations. Nevertheless, it is convenient to rest for a while to discuss two examples. They throw some light on the ways the Lenard complexes arise within the theory of integrable systems.

\bigskip
\noindent
\textbf{Example 1} Consider a bihamiltonian vector field
\[X=P_1dA=P_2dB  \]
on a bihamiltonian manifold M. If one of the Poisson bivectors, say $P_2$, is invertible, one may associate with X the recursion operator
\[K=P_1\circ P_2^{-1}.  \]
It is known, since long time, that the triple $(K,dA,X)$ satisfies the conditions
\[ Torsion(K)=0 \]
\[ Lie_X(K)=0 \]
\[ d(KdA)=0 \]
on account of the compatibility condition of the Poisson bivectors. Moreover it is known that these properties are inherited by the powers $K^i$ of K,
and that the vector fields $K^jX$, the 1-forms $K^idA$, and the 1-forms $K^jK^ldA=K^{j+l}dA$ verify accordingly the conditions defining a Lenard complex. 
Thus a Lenard complex, of  a very special type, is associated with any bihamiltonian vector field on a symplectic manifold. In this sense one may say that 
the concept of Lenard complex has its origin in the geometry of bihamiltonian manifolds. However, this setting is too restrictive. The next example shows why it 
is convenient to enlarge the old scheme of Lenard chains.

\bigskip
\noindent
\textbf{Example 2} The system of partial differential equations
\begin{align}
\frac{\partial w_2}{\partial t}=&2\frac{\partial w_1}{\partial x} \nonumber \\ 
\frac{\partial w_1}{\partial t}=&2\frac{\partial w_0}{\partial x}-w_2\frac{\partial w_2}{\partial x} \nonumber \\ 
\frac{\partial w_0}{\partial t}=&\qquad -\frac{1}{2}w_1\frac{\partial w_2}{\partial x} \nonumber 
\end{align}
is a classical example of integrable system of hydrodynamic type. It is representative of a wide class of integrable equations known as dispersionless Gelfand-Dikii equations. 
Such equations have been intensively studied in the past thirty years, or more, and accordingly it is presently known that they enjoy a lot of remarkable proprieties,
some of which are rather sophisticated and deeply concealed into the equations (see, for instance,\cite{4}). Here I want to add  a property, showing that a Lenard complex
is associated to the dispersionless Gelfand-Dikii equations. 

It is a common practice to write the above equations in matrix form, so to bring out the matrix of coefficients, 
\[V=\left( \begin{matrix}0& 2& 0\\-w_2& 0& 2 \\-\frac{1}{2}w_1& 0 &0 
\end{matrix}\right) \]
whose eigenvalues are the characteristic velocities of the system. This matrix, however, depends on the choice of the coordinates, and therefore its use is troublesome from a geometric standpoint.
To overcome this difficulty it is better to transform the matrix V into a recursion operator K, simply  by setting
\begin{align}
Kdw_2=&2dw_1 \nonumber\\
Kdw_1=&2dw_0-w_2dw_2 \nonumber\\
Kdw_0=&\qquad -\frac{1}{2}w_1dw_2 \nonumber
\end{align}
on a certain open set U of $\mathbb{R}^3$, where $(w_0,w_1,w_2)$ play the role of coordinates, and $(dw_0,dw_1,dw_2)$ is the associated canonical basis in $T^*U$.
This operator is the first character of the geometrical analysis of the Gelfand-Dikii equations. Other two characters follow from the study of K. One is related to the Nijenhuis torsion of K.
There are many ways of writing this torsion, but the most convenient is to contract the vector-valued 2-form $Torsion(K)$ with the differential of a function F, so to obtain a scalar-valued 2-form dF(Torsion(K)). In the case of Gelfand-Dikii equations, this scalar-valued 2-form has the very simple appearence
\[ dF(Torsion(K))=dw_2 \wedge dF. \]
This formula shows that the torsion of K does not vanish, and that it is completely characterized by the 1-form
\[ dA=dw_2. \] 
This form is the second interesting character of the geometric analysis of the Gelfand-Dikii equations. The third character is the vector field
\[ X=\frac{\partial}{\partial w_0} \]
which is manifestly a symmetry of K. Thus a triple of objects (K,dA,X) is naturally associated with the given equations, as in the bihamiltonian case. The main difference, however, 
is that now K has a torsion. Accordingly, one cannot  use the ``rule of powers'' to construct two chains of commuting vector fields and of exact 1-forms. 
The presence of the torsion of K obliges to make recourse to a more subtle recursive procedure. The right answer is to choose the commuting operators
\begin{align}
K_1=&Id \nonumber\\
K_2=&K \nonumber\\
K_3=&K^2+A\cdot Id. \nonumber 
\end{align}
These operators, the 1-form dA, and the vector field X satisfy all the conditions defining a Lenard complex. The existence of such a complex is another manifestation of the integrability of the Gelfand-Dikii equations.

\setcounter{equation}{0}
\renewcommand{\theequation}{\thesection.\arabic{equation}}
\newpage
\section{WDVV equations}

There is a nice and straightforward connection between Lenard complexes and WDVV equations, which I want to point out presently. It is based on the intertwining of the square of 1-forms $dA_{jl}$ with the x-coordinates of the Lenard complex.
The potentials $A_{jl}$ of the square of 1-forms are the entries of a symmetric matrix which I have called the matrix potential of the complex. I agree to write the functions $A_{jl}$ in the x-coordinates.

\bigskip
\noindent
\textbf{Proposition 1.} \textit{The matrix potential of a Lenard complex, written in the x-coordinates, is the Hessian matrix of a function F which satisfies the WDVV equations.}

\bigskip
\emph{Proof: } Let me compute the partial derivatives of the functions $A_{jl}$:
\[ \frac{\partial A_{jl}}{\partial x_m}=dA_{jl}(\frac{\partial}{\partial x^m})=dA(K_jK_lK_mX). \] 
The right-hand side is totally symmetric in the indices $(j,l,m)$, and hence the functions $A_{jl}$ are the entries of a Hessian matrix:
\[ A_{jl}=\frac{\partial ^2F}{\partial x_j \partial x_l}. \]
To show that the potential F verifies the WDVV equations, I suppose (as constantly made in the framework of the WDVV theory) that the 1-forms $dA_{jl}$, for $j=1$, are linearly independent, and therefore form a basis on $T^*M$. This assumption guarantees that the partial derivative $\frac{\partial h}{\partial x_1}$ of the Hessian matrix of the function F
\begin{equation}
\label{equaz}
h=Hessian(F)
\end{equation}
is invertible. It also guarantees that the recursion operator $K_1$ is invertible (the two facts being obviously related). 
Consider then the operators $Q_j=K_j\circ K_1^{-1}$. I notice that 
\[ Q_jdA_{1l}=dA_{jl}.\]
Hence the matrix which represents $Q_j$ on the basis $dx_j$ is
\[ Q_j=\left( \frac{\partial h}{\partial x_1}\right)^{-1} \cdot\frac{\partial h}{\partial x_j}. \] 
Since the operators $Q_j$ commute, the same do their matrices. Therefore
\begin{equation}
\label{equa}
\text{WDVV:}\qquad\frac{\partial h}{\partial x_j}\left(\frac{\partial h}{\partial x_1}\right)^{-1}\frac{\partial h}{\partial x_l}=\frac{\partial h}{\partial x_l}\left(\frac{\partial h}{\partial x_1}\right)^{-1}\frac{\partial h}{\partial x_j}.
\end{equation} 
This proves that the potential F satisfies the WDVV equations, as it was claimed.
\hfill $\square$

\bigskip
The connection between Lenard complexes and WDVV equations has many other facets which deserve attention. I briefly mention two of them, by distinguishing the case where $K_1=Id$ from the more elaborate case where $K_1\not=Id$:

\bigskip
\noindent
\emph{Case $K_1=Id$.} In this case the Lenard complex is completely characterized by the potential F, and there is a one-to-one correspondence between Lenard complexes of this type and solutions of the WDVV equations. In other words, one may easily construct a Lenard complex with $K_1=Id$ starting from any solution of the classical WDVV equations \cite{8}.

\bigskip
\noindent
\emph{Case $K_1\not=Id$.} In this case the potential F alone does not characterize completely the Lenard complex. A second function $G$ enters into the picture, which is related to the potential $F$ by a couple of interesting equations. Let  $(\lambda_1,...,\lambda_n)$ denote  the components of the vector field X in the ``x-coordinates'':
\[ X=\sum \lambda_k\frac{\partial}{\partial x^k}. \] 
Then one can prove that the matrix
\[ g=\sum \lambda_k \frac{\partial h}{\partial x_k}\]
is the Hessian matrix of a function G, and that F and G  verify the ``generalized'' WDVV equations
\begin{equation}
\label{equazi}
\frac{\partial h}{\partial x_j}g^{-1}\frac{\partial h}{\partial x_l}=\frac{\partial h}{\partial x_l}g^{-1}\frac{\partial h}{\partial x_j}.
\end{equation}

I omit the proof of these claims, but I offer an example which is pertinent to the subject of this paper. The function
\begin{equation}
\label{sol}
F(x_1x_2x_3)=\sum_{i<j}^{3}(x_i-x_j)^2\log(x_i-x_j)^2+\frac{1}{2}\sum^{3}_{i=1}x_i^2\log x_i^2
\end{equation} 
is one of the simplest solution of the WDVV equations (\ref{equa}) of the Veselov type. As we shall see in the next section, it generates a Lenard complex of the second type. Accordingly it must verify the previous equations. Indeed, one may check that the matrix  
\[ g=\sum_{i=1}^{3}\frac{1}{4}x_i\frac{\partial h}{\partial x_i} \]
is the Hessian matrix of the function
\[ G=\frac{3}{8}(x_1^2+x_2^2+x_3^2)-\frac{1}{4}(x_1x_2+x_2x_3+x_3x_2), \]
and that the functions F and G verify the generalized WDVV  equations (\ref{equazi}).The last equation is of particular interest. It entails that the function F is a solution of the ordinary WDVV equations as well. This means that certain solutions of the highly nonlinear WDVV equations can be found by solving a problem with only quadratic nonlinearities.

Another point on which I am obliged to skip is the relation with the theory of Frobenius manifolds. It is well-known that Boris Dubrovin , in the 90's , has worked out the beautiful and far reaching concept of Frobenius manifold, \cite{5},\cite{6},\cite{7} , just to give a geometrical interpretation of the WDVV equations. The concepts of Frobenius manifold and Lenard complex must, therefore, be strictly related. They provides two ways of looking at the same object from different perspectives and by using different geometrical structures. Some details of their connection, however, are still missing.

\newpage
\setcounter{equation}{0}
\section{Equivariant complexes}
The order of the recursion operators $K_1,K_2,...,K_n$ is irrelevant and these operators may be exchanged among themselves. The study of the behaviour of the Lenard complex under a change of the order of the recursion operators suggests to introduce a particular class of Lenard complexes, which are said to be equivariant with respect to the symmetric group $S_n$. They have the property that the exchange of the operators $K_j$ and $K_l$ is equivalent to the exchange of the coordinates $a_j$ and $a_l$.

Equivariant Lenard complexes are defined as follows. Let $S_n$ be the symmetric group acting on the a-coordinates $(a_1,...,a_n)$. I denote by
\[  \sigma_{jl}=(a_{j},a_l)\]
the transposition of the coordinates $a_j$ and $a_l$. I see this transposition as a map $\sigma_{jl}:M\rightarrow M$ on the manifold M. Accordingly I lift this action on the 1-forms and on the vector fields defined on M. The lifted actions are given by the pull-back $\sigma_{jl}^*$ of 1-forms, and by the push-forward $\sigma_{*jl}$ of vector fields. Examples of these actions are: 
\[ \sigma_{jl}^*(da_j)=da_l \qquad \sigma_{jl}^*(da_m)=da_m \qquad \sigma_{*jl}\left(\frac{\partial}{\partial a_j}\right)=\frac{\partial}{\partial a_l}\qquad m\not=j,l.\]
Finally, I extend the action of $S_n$  to the recursion operators according to the usual rule of tensor calculus.

\bigskip
\noindent
\textbf{Definition 4.} \textit{A Lenard complex $(K_j,dA,X)$ is said to be equivariant with respect to the action of the symmetric group $S_n$ on the a-coordinates $(a_1,...,a_n)$ if:
\begin{enumerate}
\item the vector field X and the 1-form dA are invariant
\[ \sigma_{jl}^*(dA)=dA\qquad \sigma_{*jl}(X)=X \]
\item the transposition $\sigma_{jl}:M\rightarrow M$ interchanges the recursion operators $K_j$ and $K_l$
\[ \sigma_{jl}(K_j)=K_l. \]
\end{enumerate}}
Let me work out explicitly the equivariance conditions in $dim M=3$. Let the transposition $\sigma_{jl}$ acts separately on the 1-forms of the square. This action  produces a new square. The condition of equivariance means that the new square coincides with the old square up to the exchange of the l- and j- rows and columns. Hence for equivariant complexes 
\begin{center}
\begin{tabular}{|c|c|c|c|}
\hline
$\sigma_{12}^*(dA)$&$\sigma_{12}^*(da_1)$&$\sigma_{12}^*(da_2)$&$\sigma_{12}^*(da_3)$\\
\hline
$\sigma_{12}^*(da_1)$&$\sigma_{12}^*(dP)$&$\sigma_{12}^*(dQ)$&$\sigma_{12}^*(dR)$\\
\hline
$\sigma_{12}^*(da_2)$&$\sigma_{12}^*(dQ)$&$\sigma_{12}^*(dS)$&$\sigma_{12}^*(dT)$\\
\hline
$\sigma_{12}^*(da_3)$&$\sigma_{12}^*(dR)$&$\sigma_{12}^*(dT)$&$\sigma_{12}^*(dV)$\\
\hline
\end{tabular}=
\begin{tabular}{|c|c|c|c|}
\hline
dA&$da_2$&$da_1$&$da_3$\\
\hline
$da_2$&dS&dQ&dT\\
\hline
$da_1$&dQ&dP&dR\\
\hline
$da_3$&dT&dR&$da_3$\\
\hline
\end{tabular}.
\end{center} 
This diagram shows that the 1-forms of an equivariant complex verify the symmetry constraints
\[ \sigma_{12}^*(dA)=dA \qquad  \sigma_{12}^*(dP)=dS \qquad  \sigma_{12}^*(dQ)=dQ \qquad  \sigma_{12}^*(dR)=dT.\]
One thus easily realizes that an equivariant square of 1-forms is completely identified by three 1-forms: the pivot dA, the 1-form dP on the diagonal of the square, and the 1-form dQ off the diagonal. The other 1-forms are recovered according to the relations
\[ \sigma_{12}^*(dP)=dS \qquad  \sigma_{13}^*(dP)=dV \qquad  \sigma_{23}^*(dQ)=dR \qquad  \sigma_{13}^*(dQ)=dT.\]
This procedure is clarified by the next example.

\bigskip
\noindent
\textbf{Example 3.} For reasons made clear in the next section, let me choose the initial 1-forms (dA,dP,dQ) as follows:
\begin{align}
dA=&(2a_1+a_2+a_3)da_1+(a_1+2a_2+a_3)da_2+(a_1+a_2+2a_3)da_3\nonumber\\
dQ=&-\frac{1}{4}\frac{d(a_1-a_2)}{a_1-a_2} \nonumber \\
dP=&\frac{(6a_1^2-2a_1a_2-2a_2a_3-a_2^2-a_3^2)}{4(2a_1+a_2+a_3)(a_1-a_2)(a_1-a_3)}da_1-\frac{(2a_2+a_1+a_3)}{4(2a_1+a_2+a_3)(a_1-a_2)}da_2+ \nonumber\\
&\quad-\frac{(2a_3+a_1+a_2)}{4(2a_1+a_2+a_3)(a_1-a_3)}da_3\nonumber
\end{align}
By the action of the symmetric group $S_3$ on the coordinates $(a_1,a_2,a_3)$, I first construct the orbit passing through dQ. I  obtain 
\[ dQ=-\frac{1}{4}\frac{d(a_1-a_2)}{a_1-a_2} \quad dR=-\frac{1}{4}\frac{d(a_1-a_3)}{a_1-a_3} \quad dT=-\frac{1}{4}\frac{d(a_2-a_3)}{a_2-a_3}.\]
Similarly from dP, I obtain
\begin{align}
dP&=\frac{(6a_1^2-2a_1a_2-2a_2a_3-a_2^2-a_3^2)}{4(2a_1+a_2+a_3)(a_1-a_2)(a_1-a_3)}da_1-\frac{(2a_2+a_1+a_3)}{4(2a_1+a_2+a_3)(a_1-a_2)}da_2+ \nonumber\\
&\quad-\frac{(2a_3+a_1+a_2)}{4(2a_1+a_2+a_3)(a_1-a_3)}da_3\nonumber\\
dS&=-\frac{(2a_1+a_2+a_3)}{4(2a_2+a_1+a_3)(a_2-a_1)}da_1+\frac{(6a_2^2-2a_2a_3-2a_2a_1-a_3^2-a_1^2)}{4(2a_2+a_1+a_3)(a_2-a_1)(a_2-a_3)}da_2+\nonumber\\
&\quad-\frac{(2a_3+a_2+a_1)}{4(2a_2+a_1+a_3)(a_2-a_3)}da_3\nonumber\\
dV&=-\frac{(2a_1+a_2+a_3)}{4(2a_3+a_2+a_1)(a_3-a_1)}da_1-\frac{(2a_2+a_1+a_3)}{4(2a_3+a_2+a_1)(a_3-a_2)}da_2+\nonumber\\
&\quad+\frac{(6a_3^2-2a_1a_3-2a_2a_3-a_1^2-a_3^2)}{4(2a_3+a_2+a_1)(a_3-a_1)(a_3-a_2)}da_3.\nonumber
\end{align}
In this way the square is completed  by equivariance.
Next I use the square of 1-forms to define the recursion operators $(K_1,K_2,K_3)$ according to
\begin{align}
K_1(da_1)=dP\qquad K_1(da_2)=dQ\qquad K_1(da_3)=dR \nonumber \\
K_2(da_1)=dQ\qquad K_2(da_2)=dS\qquad K_2(da_3)=dT \nonumber \\
K_3(da_1)=dR\qquad K_3(da_2)=dT\qquad K_3(da_3)=dV. \nonumber
\end{align}
To complete the Lenard complex, we are still missing the vector field X. Again for reasons explained in the next section, let me choose the vector field
\[ X=a_1\frac{\partial}{\partial a_1}+a_2\frac{\partial}{\partial a_2}+a_3\frac{\partial}{\partial a_3}. \]
The iterated vector fields $(K_1X,K_2X,K_3X)$ are
\begin{align}
X_1&=\frac{1}{4}\left(+3\frac{\partial}{\partial a_1}-\frac{\partial}{\partial a_2}-\frac{\partial}{\partial a_3}\right)\nonumber\\
X_2&=\frac{1}{4}\left(-\frac{\partial}{\partial a_1}+3\frac{\partial}{\partial a_2}-\frac{\partial}{\partial a_3}\right)\nonumber\\
X_3&=\frac{1}{4}\left(-\frac{\partial}{\partial a_1}-\frac{\partial}{\partial a_2}+3\frac{\partial}{\partial a_3}\right)\nonumber
\end{align}
They  commute, and therefore they define the x-coordinates. Such coordinates are related to the previous a-coordinates by the relations
\begin{align}
x_1=&2a_1+a_2+a_3\nonumber\\
x_2=&2a_2+a_3+a_1\nonumber\\
x_3=&2a_3+a_1+a_2.\nonumber
\end{align}
One may check that the complex presently constructed verifies the condition imposed on a Lenard complex. Accordingly, by Proposition 1, it defines implicitly a solution F of the WDVV equations. This solution can be worked out by writing the 1-forms of the square in the x-coordinates. One may thus check the identities $dQ=dF_{12}$, $dR=dF_{13}$, $dT=dF_{23}$, $dP=dF_{11}$, $dS=dF_{22}$, $dV=dF_{23}$, where $F_{jk}$ are the second partial derivatives of the function
\[ F(x_1x_2x_3)=\frac{1}{16}\sum_{i<j}^{3}(x_i-x_j)^2\log(x_i-x_j)^2+\frac{1}{16}\sum^{3}_{i=1}x_i^2\log x_i^2\]
with respect to the coordinates $x_j$ and $x_k$.
This computation shows that the simplest solution of the Veselov type of the WDVV equations is the potential of an equivariant Lenard complex. 
\newpage
\setcounter{equation}{0}
\section{Veselov systems}
The requirement of equivariance, by imposing the symmetry constraints, significantly reduces the number of free parameters which one has at his disposal to construct a Lenard complex. At the same time, however, it also reduces the number of independent conditions which must be satisfied by the parameters. The purpose of this section is to show that the balance between the reduction of the number of degrees of freedom and the reduction of the number of conditions is favorable. This circumstance will allow me to construct a family of equivariant Lenard complexes on $\mathbb{R}^3$, depending on one arbitrary parameter.

The main idea is that the equations defining a Lenard complex may be solved in a purely algebraic way in the class of equivariant differential forms of logarithmic type, that is in the class of 1-forms of the type
\[ dQ=\sum^{m}_{k=1}\sigma_k\frac{dB_k}{B_k}, \]
where the $\sigma_k$'s are parameters, and the $B_k$'s are functions to be suitably chosen. It would be quite instructive to explain the reason for the choice of this class of forms through a closer inspection of the structure of the conditions defining a Lenard complex. For brevity, however, I will present only the final result.

Let 
\begin{equation}
A=\frac{1}{2}\alpha(a_1^2+a_2^2+a_3^2)+\beta(a_1a_2+a_2a_3+a_3a_2) 
\end{equation}
be the most general quadratic polynomial on $\mathbb{R}^3$ which is invariant under the action of the symmetric group $S_3$. I assume that
\[ (\alpha-\beta)^2(2\beta+\alpha)\not=0, \]
so that the Hessian matrix of the function A is invertible. I notice that the partial derivatives of A
\begin{align}
A_1=&\alpha a_1+\beta (a_2+a_3)\nonumber \\
A_2=&\alpha a_2+\beta (a_3+a_1)\nonumber \\
A_3=&\alpha a_3+\beta (a_1+a_2)\nonumber 
\end{align}
provide, in this case, a new system of coordinates on $\mathbb{R}^3$. It is obvious that the transposition $\sigma_{jl}$ acts on the new coordinates exactly as it acts on the a-coordinates. In the new coordinates I define the pair of 1-forms
\begin{equation}
dQ=\sum_{k=1}^m \sigma_k\left( \frac{d(A_1+\eta_kA_2)}{A_1+\eta_kA_2}+\frac{d(A_2+\eta_kA_1)}{A_2+\eta_kA_1}\right)
\end{equation}
and
\begin{align}
dP=\sigma_0\frac{dA_1}{A_1}&+\sum_{k=1}^m\frac{\sigma_k}{\eta_k}\left(\frac{d(A_1+\eta_kA_2)}{A_1+\eta_kA_2}+\frac{d(A_1+\eta_kA_3)}{A_1+\eta_kA_3}\right)\\
\qquad&+\sum_{k=1}^m \sigma_k\eta_k\left(\frac{d(A_2+\eta_kA_1)}{A_2+\eta_kA_1}+\frac{d(A_3+\eta_kA_1)}{A_3+\eta_kA_1}\right), \nonumber
\end{align}
where $(\sigma_0,\sigma_k,\eta_k)$ are $(2m+1)$ parameters. They are assumed to obey the constraints
\begin{equation}
\label{c}
 2\sum^{m}_{k=1}\sigma_k=\frac{\beta}{(\beta-\alpha)(2\beta+\alpha)}
 \end{equation}
\begin{equation}
\label{d}
 \sigma_0+2\sum^{m}_{k=1}\left(\frac{\sigma_k}{\eta_k}+\sigma_k\eta_k\right)=\frac{\alpha+\beta}{(\alpha-\beta)(2\beta+\alpha)} .\end{equation}
The above 1-forms manifestly satisfy the symmetry constraints
\[ \sigma_{12}^*(dQ)=dQ \qquad \sigma_{23}^*(dP)=dP, \]
and accordingly they can be used to construct an equivariant square of 1-forms, following the procedure illustrated in the previous section. I denote, as before, by
\[ dR=\sigma_{23}^*(dQ) \]
\[ dT=\sigma_{13}^*(dQ) \]
the elements of the orbit of the symmetric group passing through dQ, and by
\[ dS=\sigma_{12}^*(dP) \]
\[ dV=\sigma_{13}^*(dP) \]
those of the orbit passing through dP. Then I use the six 1-forms (dQ,dR,dT,dP,dS,dV) to define the recursion operators $(K_1,K_2,K_3)$ according to the standard relations (on the ``a-coordinates'')
\begin{align}
K_1(da_1)=dP\qquad K_2(da_1)=dQ\qquad K_3(da_1)=dR \nonumber \\
K_1(da_2)=dQ\qquad K_2(da_2)=dS\qquad K_3(da_2)=dT \nonumber \\
K_1(da_3)=dR\qquad K_2(da_3)=dT\qquad K_3(da_3)=dV. \nonumber
\end{align}
Finally, I consider the vector field
\[X=A_1\frac{\partial}{\partial A_1}+A_2\frac{\partial}{\partial A_2}+A_3\frac{\partial}{\partial A_3},\]
and I set
\[X_1=\frac{\partial}{\partial A_1}\qquad X_2=\frac{\partial}{\partial A_2} \qquad X_3=\frac{\partial}{\partial A_3}.\]

\bigskip
\noindent
\textbf{Proposition 2.}\textit{ The tensor fields $(K_j,dA,X)$, given above, define an equivariant Lenard complex if the parameters $(\sigma_k,\eta_k)$ are chosen in such a way that the 1-form $K_3dQ$ satisfies the symmetry condition 
\begin{equation}
\label{equal}
\sigma_{23}^*(K_3dQ)=K_3dQ
\end{equation}
under the transposition of the coordinates $A_2$ and $A_3$. Hence, by Proposition1, each solution of this condition implicitly defines a solution of the WDVV equations on $\mathbb{R}^3$.}

\bigskip 
\noindent
\emph{Proof. } I have to show that the above tensor fields verify the four conditions
\renewcommand{\theenumi}{\Roman{enumi}}%
\begin{enumerate}
\item $K_j(dA)=da_j$
\item $K_j(X)=X_j$
\item $K_jK_l(dA)=dA_{jl}$
\item $K_jK_l-K_lK_j=0.$
\end{enumerate}
The first condition says that $(K_j,dA)$ define a chain of exact 1-forms. The second condition says that $(K_j,X)$ define a chain of commuting vector fields. The third condition says that $(K_j,dA)$ define  a square of exact 1-forms. The fourth condition, finally, implies that this square is symmetric. Without this condition the matrix potentials of the Lenard complex cannot be the Hessian matrix of a function F which satisfies the WDVV equations. According to the splitting of the conditions in four groups, I divide the proof in four parts.

\bigskip
\noindent
\emph{1.Chain of 1-forms.} I consider first the condition $K_1(dA)=da_1$. By expanding dA on the basis $da_j$, I write this condition in the equivalent form
\[ A_1dP+A_2dQ+A_3dR=da_1 \] 
on account of the definition of $K_1$. Inserting into this equation the definitions of the 1-forms dP, dQ, and dR, one realizes that the denominators cancel out, and that the condition simplifies into the \textit{linear} equation
\begin{align}
\sigma_0dA_1&+\sum_{k}\frac{\sigma_k}{\eta_k}\left[d(A_1+\eta_kA_2)+d(A_1+\eta_kA_3)\right]+\nonumber\\
\qquad&+\sum_{k}\sigma_k\left[d(A_2+\eta_kA_1)+d(A_3+\eta_kA_1)\right]=da_1. \nonumber
\end{align}
This cancellation is the deep reason for the specific choice of the 1-forms dP and dQ. The equation is now immediately integrated, and its validity follows from $(\ref{c})$ and $(\ref{d})$. The other conditions $K_2(dA)=da_2$ and $K_3(dA)=da_3$ follow by equivariance.

\bigskip
\noindent
\emph{2.Chain of vector fields.} Let me evaluate the vector field $K_1(X)$ on the basis $da_j$. I find:
\begin{align}
da_1(K_1X)&=dP(X) \nonumber\\
\qquad&=\sigma_0+2\sum_k\left(\frac{\sigma_k}{\eta_k}+\sigma_k\eta_k\right)\nonumber\\
\qquad&=\frac{(\alpha+\beta)}{(\alpha-\beta)(2\beta+\alpha)}\nonumber\\
\qquad&=da_1(X_1).\nonumber
\end{align}
Similarly, I find 
\[ da_2(K_1X)=2\sum\sigma_k=\frac{\beta}{(\beta-\alpha)(2\beta+\alpha)}=da_2(X_1) \]
\[ da_3(K_1X)=2\sum\sigma_k=\frac{\beta}{(\beta-\alpha)(2\beta+\alpha)}=da_3(X_1). \]
Together, these identities show that $K_1(X)=X_1$, as it was required to prove. The other two conditions $K_2(X)=X_2$ and $K_3(X)=X_3$ follow by equivariance.

\bigskip
\noindent
\emph{3.Commutativity of the recursion operators.} I evaluate the commutator $[K_2,K_1]$ on the basis $da_j$. I obtain:
\begin{align}
[K_2,K_1](da_1)=K_2dP-K_1dQ\nonumber \\
[K_2,K_1](da_2)=K_2dQ-K_1dS\nonumber\\
[K_2,K_1](da_3)=K_2dR-K_1dT\nonumber
\end{align}
owing to definition of the recursion operators. I notice that the second relation follows, by equivariance, from the first relation. Similarly, I notice that the conditions for $[K_3,K_1]$ and $[K_3,K_2]$ follow from the previous ones, always by equivariance. Thus I am left with the pair of commutativity conditions 
\begin{align}
\label{b}
 K_2dP=K_1dQ  \\
  K_2dR=K_1dT. 
  \label{a}
\end{align}

I notice that the second condition follows from the assumption $(\ref{equal})$.
Indeed, by equivariance
\[ \sigma_{23}(K_3dQ)=K_2\sigma_{23}(dQ)=K_2dR, \]
and thus the assumption $(\ref{equal})$ implies
\[ K_2dR=K_3dQ. \]
The relation $(\ref{a})$ then follows by equivariance: 
\begin{align}
K_2dR&=\sigma_{13}(K_2dR) \nonumber\\
\qquad&=\sigma_{13}(K_3dQ)\nonumber\\
\qquad&=K_1\sigma_{13}(dQ)\nonumber\\
\qquad&=K_1dT.\nonumber
\end{align}

As a final step, I notice that the condition $(\ref{b})$ follows from $(\ref{equal})$ as well. Indeed the condition is equivalent to 
\[ A_1K_2dP=A_1K_1dQ. \]
Since
\[ A_1dP+A_2dQ+A_3dR=da_1, \]
I get
\begin{align}
A_1K_2dP&=-(A_2K_2dQ+A_3K_2dR)+K_2da_1 \nonumber\\
\qquad&=-(A_2K_2dQ+A_3K_3dQ)+dQ. \nonumber
\end{align}
So the task is to prove that
\[ A_1K_1(dQ)+A_2K_2(dQ)+A_3K_3(dQ)=dQ. \]
This relation is true since
\[ A_1K_1+A_2K_2+A_3K_3=Id. \]
Indeed
\[ (A_1K_1+A_2K_2+A_3K_3)da_1=A_1dP+A_2dQ+A_3dR=da_1\]
and similarly for $a_2$ and $a_3$. It remains thus proved that the recursion operators commute on account of assumption $(\ref{equal})$.

\bigskip
\noindent
\emph{4.The square of 1-forms.} The relations $K_jK_l(dA)=dA_{jl}$ are a simple consequence of the definition of the recursion operators.
\hfill $\square$

\bigskip
The possibility of constructing equivariant Lenard complexes on $\mathbb{R}^3$ (and, probably, on $\mathbb{R}^n$) depends therefore on our ability of solving the constraint $(\ref{equal})$. It is a very restrictive constraint, since the unknowns are parameters, while the constraint depends explicitly (and in a rather nonlinear and complicated way) on the coordinates. In a sense it is surprising that it admits solutions. I describe presently a solution of the symmetry constraint $(\ref{equal})$ in the class of forms of logarithmic type considered in this section.

The solution concerns the case $m=2$, where the 1-forms of the Lenard complex depend on seven parameters $(\alpha,\beta;\sigma_0,\sigma_1,\sigma_2;\eta_1,\eta_2)$. Two of them, $\sigma_0$ and $\sigma_1$, are fixed by the conditions $(\ref{c}),(\ref{d})$. The other parameters must be chosen so to verify the symmetry constraint $(\ref{equal})$ at a generic point $(A_1,A_2,A_3)$. It turns out that the choice of the parameters $(\eta_1,\eta_2)$ is essential to this end. Indeed if one sets
\begin{equation}
\label{cond}
\eta_1=+1 \qquad\eta_2=-1,
\end{equation}
one discovers that the 1-form $\sigma_{23}^*(K_3dQ)-K_3dQ$ takes the nice \textit{splitted} form
\[ \sigma_{23}^*(K_3dQ)-K_3dQ=\Phi(\alpha,\beta,\sigma_2)\Psi(A_1,A_2,A_3)\left(\frac{dA_3}{A_3}-\frac{dA_2}{A_2}\right), \]
where
\begin{small}
\[ 
\Psi(A_1,A_2,A_3)=\frac{A_1A_2+A_2A_3+A_3A_1}{(A_1+A_2)(A_2+A_3)(A_3+A_1)} \]
\end{small}
and
\begin{small}
\[\Phi(\alpha,\beta,\sigma_2)=\frac{2\beta(2\alpha\beta\sigma_2+2\alpha^2\sigma_2-4\beta^2\sigma_2+\beta)(8\alpha\beta\sigma_2+8\alpha^2\sigma_2-16\beta^2\sigma_2+\alpha+3\beta)}{(\alpha-\beta)^2(2\beta+\alpha)^2}.\] \end{small}
Hence the symmetry constraint $(\ref{equal})$ is satisfied by any choice of the parameters $(\alpha,\beta,\sigma_2)$ such that 
\begin{equation}
\label{condition3}
\Phi(\alpha,\beta,\sigma_2)=0.
\end{equation} 
There are three possible choices, each leaving two parameters free. However, one parameter may always be rescaled. So the final Lenard complex depends effectively on a single free parameter.

The case $m=2$, just discussed, is basic. I believe that the general case with $m>2$ will actually reduce  to this case . Indeed the symmetry constraint $(\ref{equal})$ is so strong to imply always, I believe,  the condition $\eta_k=\pm 1$. For this reason, it seems to me plausible that there should not be other equivariant complexes on $\mathbb{R}^3$, beyond the complex previously exhibited. This one-parameter family of equivariant Lenard complexes is, however, sufficiently general to encompass the Veselov solutions of type $A_3$ and $B_3$ at least.
\section{Concluding remarks}
In this paper I have tried to argue that the geometric standpoint of the Lenard complexes is a convenient bridge to arrive to the solution of the WDVV equations. I have also shown that the notion of equivariance allows to reduce the construction of an equivariant Lenard complex on $\mathbb{R}^3$ to an \textit{algebraic} problem, namely to solve the symmetry constraint 
\[ \sigma_{23}(K_3dQ)=K_3dQ \]
in the class of 1-forms of logarithmic type.
Finally, I have shown that this constraint always admits, on $\mathbb{R}^3$, a one-parametric family of solutions. Each of them induces a solution of the WDVV equations. The question of how many of the Veselov solutions are contained in this one-parameter family of equivariant complexes is still open. What remains unclear, so far, it is if the Lenard complexes associated with the Veselov solutions, according to Proposition 1, are all equivariant or not. 

\bigskip
\noindent
\emph{\textbf{Acknowledgements}}
I am particularly grateful to Boris Konopelchenko for his friendly support during this research  on the WDVV equations. I wish to thank also Paolo Lorenzoni and Andrea Raimondo for useful discussions on the Veselov solutions of type $A_3$ and $B_3$.  

\end{document}